%% file: main.tex
\documentclass{article}
\usepackage{spconf}

\title{{\fontsize{10.97pt}{0pt}\selectfont What Did Your Adversary Believe? \\ 
Optimal Filtering and Smoothing in Counter-Adversarial Autonomous Systems}}
\name{Robert Mattila$^{\star}$, In\^{e}s Louren\c{c}o$^{\star}$, Vikram
    Krishnamurthy$^{\dagger}$, Cristian R.
    Rojas$^{\star}$ and Bo Wahlberg$^{\star}$ \thanks{This work was supported by the Wallenberg AI, Autonomous
        Systems and Software Program (WASP), the Swedish Research Council (2016-06079) and
the U.S. Air Force Office of Scientific Research (FA9550-18-1-0007).}}
\address{$^{\star}$KTH Royal Institute of Technology \hspace{2cm} $^{\dagger}$Cornell University\\
    \hspace{2.1cm} Decision and Control Systems \hspace{1cm} Electrical and Computer Engineering\\
    \hspace{0.3cm} {\tt\footnotesize\{rmattila,ineslo,crro,bo\}@kth.se \hspace{2.3cm}
vikramk@cornell.edu}}

\newif\ifIFAC                                                                  
\IFACfalse

\input{pre_rob.tex}

\usepackage{balance}

\newcommand{\bv}[1] {{\beta_{#1}}} 
\newcommand{\bvr}[1] {{\beta^\natural_{#1}}} 
\newcommand{\fv}[1] {{\alpha_{#1}}} 
\newcommand{\action} {a} 
\newcommand{\bset} {\Pi} 
\newcommand{\rpolicy} {G} 

\allowdisplaybreaks

\begin{document}
\ninept


\maketitle

\begin{abstract}

    We consider fixed-interval smoothing problems for counter-adversarial autonomous
    systems. An adversary deploys an autonomous filtering and control system that
    \emph{i)} measures our current state via a noisy sensor, \emph{ii)} computes
    a posterior estimate (belief) and \emph{iii)} takes an action that we can observe.
    Based on such observed actions and our knowledge of our state sequence, we aim to
    estimate the adversary's past and current beliefs -- this forms a foundation for
    predicting, and counteracting against, future actions. We derive the optimal smoother
    for the adversary's beliefs (we treat the problem in a Bayesian framework). Moreover,
    we demonstrate how the smoother can be computed for discrete systems even though the
    corresponding backward variables do not admit a finite-dimensional characterization.
    Finally, we illustrate our results in numerical simulations.

\end{abstract}


\section{Introduction}

During the last decade, the importance of defense against cyber-adversarial and autonomous
treats has been highlighted on numerous occasions -- e.g., \cite{kuptel_counter_2017,
    barni_coping_2013, farwell_stuxnet_2011}. In this paper, we consider the design of
    \emph{counter-adversarial autonomous} (CAA) systems.

In a CAA system, an adversary employs an autonomous filtering and control system that
estimates our state and takes actions based on its control policy
\cite{kuptel_counter_2017, krishnamurthy_how_2019}. Mathematically, it can
be seen as a game between two players (\emph{us} and the \emph{adversary}), with the
following dynamics:
\begin{subequations}
\begin{align}
    \text{us:}\quad x_k &\sim P_{x_{k-1}, x} = p(x | x_{k-1}), \;\; x_0 \sim \pi_0, \label{eq:game_model1} \\
    \text{adversary:}\quad y_k &\sim B_{x_k, y} = p(y | x_k), \\
    \text{adversary:}\quad \pi_k &= T(\pi_{k-1}, y_k), & \label{eq:game_model3} \\
    \text{adversary \& us:}\quad \action_k &\sim \rpolicy_{\pi_k, \action} = p(\action | \pi_k), & \label{eq:game_model4}
\end{align}
\end{subequations}
where by $\sim$ we mean ``distributed according to'' and $k$ denotes discrete time. To be
more specific, the model \eref{eq:game_model1}-\eref{eq:game_model4} should be interpreted
as follows:
\begin{itemize}
    \item $x_k \in \mathcal{X}$ is \emph{our} Markovian state that
        evolves according to a transition kernel $P$. Its initial distribution is $\pi_0$.

    \item $y_k \in \mathcal{Y}$ is \emph{the adversary's} observation of our current
        state. The observation is sampled according to the observation likelihoods $B$.

    \item The map $T$ is the classical Bayesian filter (e.g.,
        \cite{krishnamurthy_partially_2016, cappe_inference_2005}):
        \begin{equation}
            \{T(\pi, y)\}(x) = \frac{B_{x, y} \int_\mathcal{X} P_{\zeta, x} \pi(\zeta)
            d\zeta}{\int_\mathcal{X} B_{x, y} \int_\mathcal{X} P_{\zeta, x} \pi(\zeta)
        d\zeta dx},
            \label{eq:bayesian_filter}
        \end{equation}
        that computes the belief $\pi_k(x) = p(x_k = x \, | \, y_1, \dots, y_k)$
        recursively. This update is performed \emph{by the adversary} who is trying to
        estimate our state $x_k$.
    
    \item The adversary selects an action based on its current belief $\pi_k$, and
        $\action_k \in \mathcal{A}$ is \emph{our} measurement of it. Note that $\rpolicy$
        allows for a randomized policy and/or a noisy observation of the action.

\end{itemize}
The central question that we pose in this paper is:
\begin{quote}
    
    \emph{Given what is known to us (i.e., the state sequence $x_0, \dots, x_k$ and the
    observed actions $\action_1, \dots, \action_k$), how  to estimate the adversary's
beliefs $\pi_1, \dots, \pi_k$?}

\end{quote} 
Reconstructing the adversary's beliefs forms a foundation for analyzing the behaviour of
the adversary; it provides a way of predicting and, hence also taking appropriate
counter-actions against, future actions. Moreover, it leads to other important
questions: How accurate are the adversary's sensors? How should our state sequence
(transition kernel) be designed so as to estimate the adversary's sensors as accurately as
possible and/or to confuse it? These questions have practical implications in, not only
electronic warfare and cyber-physical security, but also in, e.g., radar calibration and
interactive learning \cite{krishnamurthy_how_2019}.  \looseness=-1

In summary, the main results of this paper are:
\begin{itemize}

    \item We derive a recursion for the optimal smoother for estimating the adversary's
        beliefs given the state sequence and its observed actions;

    \item For discrete CAA systems, we provide a finite algorithm for computing the
        fixed-interval smoother -- even though the backward variables do not admit
        a finite-dimensional characterization;

    \item We illustrate and evaluate the performance of the optimal smoother in numerical
        simulations.

\end{itemize}


\subsection{Related Work}
\label{sec:related_work}

Inverse estimation and control problems have a long history in signal processing and
automatic control. For example, Kalman studied the inverse optimal control problem already
in 1964 \cite{kalman_when_1964} (aiming to determine for what cost criteria a given
control policy is optimal). More recently, inverse problems in image signal processing
(e.g., denoising and medical image reconstruction) have received attention
\cite{mccann_convolutional_2017}.

The problems considered in the present paper are a type of inverse filtering problem
\cite{mattila_inverse_2017, mattila_inverse_2018, krishnamurthy_how_2019}. Algebraic
solutions to the problem of reconstructing sensor parameters given posteriors from hidden
Markov models and linear Gaussian state-space models have been proposed in
\cite{mattila_inverse_2017} and \cite{mattila_inverse_2018}, respectively. In a CAA-system
scenario, these works assume direct access to the adversary's beliefs and use them to
infer its sensor's specifications -- more realistically, the adversary would reveal only
actions based on its beliefs.  

To deal with this assumption, \cite{ mattila_estimating_2019} considers sequential
stochastic decision problems and determines the set of beliefs that a rational adversary
could have held given an observed action. The proposed solution is based on inverse
optimization techniques and, essentially, inverts relation \eref{eq:game_model4}. The work
is oblivious to the process generating the adversary's beliefs, which, in a CAA system, is
the model \eref{eq:game_model1}-\eref{eq:game_model3}. In comparison, in the present
paper, we take the full generative model into account and compute the Bayesian posterior
of the adversary's beliefs. Due to being model-based, we can rule out beliefs that are
feasible with respect to the adversary's policy, but not with respect to the dynamics of
the game. Moreover, we obtain a full probabilistic characterization of how likely
different beliefs are.

We build on and extend the recent work \cite{krishnamurthy_how_2019} (and its
journal-preprint \cite{krishnamurthy_how_2019_arxiv}) in which a Bayesian framework for
inverse filtering in CAA systems was proposed. In particular,
\cite{krishnamurthy_how_2019} derived the optimal inverse filter (which computes the
posterior over the adversary's \emph{current} belief given a state-sequence and actions).
We employ this result to derive the optimal smoother (which computes the posteriors over
\emph{all the adversary's past beliefs}).  This yields more accurate estimates (in
a mean-squared error sense), since more information is included in the smoother than the
filter \cite{anderson_optimal_1979}.

The organization of the paper is as follows. Preliminaries and formal problem definitions
are given in \sref{sec:prelim}. In \sref{sec:inv_filt}, we show how the smoother's forward
variables can be computed. The paper's main contributions are in \sref{sec:inv_smooth}
where the optimal smoother for estimating the adversary's beliefs is derived. There, we
also specialize to discrete CAA systems. Finally, we evaluate the proposed algorithms in
numerical simulations in \sref{sec:numerical}.

\section{Preliminaries and Problem Formulation}
\label{sec:prelim}

In this section, we detail our notation and how the model
\eref{eq:game_model1}-\eref{eq:game_model4} instantiates itself for discrete
\emph{counter-adversarial autonomous} (CAA) systems.  We also provide formal statements of
the problems we treat in this paper.

\subsection{Notation}

All vectors are column vectors unless transposed. The vector of all ones is denoted
$\ones$ and the $i$th Cartesian basis vector $e_i$. The element at row $i$ and column $j$
of a matrix is $[\cdot]_{ij}$, and the element at position $i$ of a vector is $[\cdot]_i$.
The vector operator $\diag{\cdot}:\mathbb{R}^n \rightarrow \mathbb{R}^{n\times n}$ gives
the matrix where the vector has been put on the diagonal, and all other elements are zero.
The indicator function $\ind{\cdot}$ takes the value 1 if the expression $\cdot$ is
fulfilled and 0 otherwise.  We employ $\xi_{0:k}$ as shorthand for the sequence $\xi_0,
\dots, \xi_k$, and define $\xi_{k+1:k} = \varnothing$.

\subsection{Discrete CAA Systems}
\label{sec:discrete_caa}

In discrete CAA systems, the model \eref{eq:game_model1}-\eref{eq:game_model4} takes the
form:
\begin{itemize}
    \item Our state $x_k \in \mathcal{X} = \{1, \dots, X\}$ is finite. It evolves
        according to a transition probability matrix $P$ with elements $[P]_{ij}
        = \Pr{x_{k+1} = j | x_k = i},$ where $1 \leq i,j \leq X$.

    \item The adversary's observation $y_k \in \mathcal{Y} = \{1, \dots, Y\}$ is also
        finite. It is sampled according to an observation probability matrix $B$ with elements
        $[B]_{ij} = \Pr{y_k = j | x_k = i},$ where $1 \leq i \leq X$ and $1 \leq j \leq
        Y$. 

    \item Note that $x_k$ and $y_k$ define a discrete-time \emph{hidden Markov model}
        (HMM). Hence, the Bayesian filter $T$ employed by the adversary is the HMM filter (e.g.,
        \cite{krishnamurthy_partially_2016, cappe_inference_2005}):
        \begin{equation}
            \pi_{k} = T(\pi_{k-1}, y_k) = \frac{B_{y_{k}} P^T \pi_{k-1}}{\ones^T B_{y_{k}} P^T \pi_{k-1}},
            \label{eq:hmm_filter}
        \end{equation}
        where the belief $\pi_k \in \Rb^X$ is a a non-negative vector $[\pi_{k}]_i = \Pr{x_k
= i | y_{1:k}}$, and $B_{y} = \diag{B e_y} \in \Rb^{X\times X}$ is a diagonal matrix of 
the $y$th column of the observation matrix $B$. 

    \item $\action_k \in \mathcal{A} = \{1, \dots, A\}$ is our measurement of the
        adversary's action based on its current belief $\pi_k$. 

\end{itemize}

\subsection{Inverse Filtering Problems for CAA Systems}

Formally, the problems we consider in this paper are:
\begin{problem}

    Consider the CAA system \eref{eq:game_model1}-\eref{eq:game_model4}. Suppose
    probability distributions $P$, $\rpolicy$ and $B$ are known, as well as the initial
    belief $\pi_0$. Given a state sequence $x_0, \dots, x_N$ and observed actions $a_1,
    \dots, a_N$, estimate the adversary's belief $\pi_k$ by computing:
    \begin{itemize}
        \item[a)] the \emph{filter} posterior $p(\pi_k | \action_{1:k}, x_{0:k})$;
        \item[b)] the \emph{smoother} posterior $p(\pi_k | \action_{1:N}, x_{0:N})$,
    \end{itemize}
    where $k \leq N$.

    \label{pr:inverse_filter_caa}
\end{problem}

Recall that from a practical point of view, estimating the adversary's belief allows us to
predict (in a Bayesian sense) future actions of the adversary. In the design of CAA
systems, this facilitates taking effective measures against such actions. 

\begin{remark} 

    In terms of our assumptions; since we control the state sequence via $P$, we can argue
    for perfect knowledge of it. The adversary's policy and $\rpolicy$ can be estimated by
    reasoning about what is rational of the adversary's to do in different scenarios.
    Knowledge of the adversary's sensor $B$ is a stronger assumption. However, the
    solution to \pref{pr:inverse_filter_caa} (under this assumption) is \emph{part of} the
    solution proposed in \cite{krishnamurthy_how_2019} on how to estimate the adversary's
    sensor $B$ in a maximum-likelihood sense (based on its observed actions and the state
    sequence) -- hence, \pref{pr:inverse_filter_caa} can be seen as a means to an end.

\end{remark}

\section{Optimal Inverse Filter \\ for Estimating the Current Belief}
\label{sec:inv_filt}

In this section, we present the solution of the inverse filtering problem for CAA systems
that was proposed in \cite{krishnamurthy_how_2019} -- it estimates the adversary's beliefs
by a filtering recursion. This is a crucial component of the optimal smoother we derive in
the next section.

\subsection{General Inverse Filter}

Under the model defined in \eref{eq:game_model1}-\eref{eq:game_model4}, the aim of
\pref{pr:inverse_filter_caa}a is to compute the posterior of the adversary's current belief given
knowledge of our state sequence and its recorded actions up to the present time:
\begin{equation}
    \fv{k}(\pi) \eqdef p(\pi_k = \pi | \action_{1:k}, x_{0:k}).
    \label{eq:fv_k_def}
\end{equation}
Before providing the details, note that $\fv{k}(\cdot)$ is a density over $\bset$, where
$\bset$ is the set of all beliefs from \eref{eq:bayesian_filter}. In particular, if the
state-space $\mathcal{X}$ is continuous, then $\bset$ is a function space comprising the
space of density functions; if $\mathcal{X}$ is finite, then $\bset$ is the unit
$(X-1)$-dimensional simplex (corresponding to $X$-dimensional probability vectors).

The key result in \cite{krishnamurthy_how_2019} is a recursive algorithm for computing the
posterior $\fv{k}(\pi)$:
\begin{theorem}[\cite{krishnamurthy_how_2019}]
    The posterior $\fv{k}$ in \eref{eq:fv_k_def} satisfies the following filtering
    recursion initialized by a prior random measure $\fv{0} = p(\pi_0)$:
    \begin{equation}
        \fv{{k+1}}(\pi) = \frac{\rpolicy_{\pi, \action_{k+1}} \int_\bset B_{x_{k+1},
        y_{\pi_k, \pi}} \fv{k}(\pi_k) d\pi_k}{\int_\bset \rpolicy_{\pi, \action_{k+1}} \int_\bset B_{x_{k+1},
        y_{\pi_k, \pi}} \fv{k}(\pi_k) d\pi_k d\pi}.
        \label{eq:alpha_recursion}
    \end{equation}
    Here, $y_{\pi_k, \pi}$ is the observation such that $\pi = T(\pi_k, y)$ where $T$ is
    the adversary's filter \eref{eq:game_model3}. The conditional mean estimate of the
    belief is $\E{\pi_{k+1} | \action_{1:k}, x_{0:k}} = \int_{\bset} \pi \fv{{k+1}}(\pi)
    d\pi$.

    \label{thrm:inverse_filter}
\end{theorem}

We refer to the update \eref{eq:alpha_recursion} as the \emph{optimal inverse filter}. For
simplicity, we assume the initial prior to be a Dirac-delta function placed in $\pi_0$,
i.e., $\fv{0}(\pi) = \delta(\pi - \pi_0)$.

\subsection{Inverse Filter for Discrete CAA Systems}

Depending on the characteristics of the adversary's belief space $\bset$, the integrals in
\eref{eq:alpha_recursion} can be tractable or not to compute. A few special cases admit
finite-dimensional characterizations of the optimal inverse filter: discrete CAA systems
and linear-Gaussian CAA systems \cite{krishnamurthy_how_2019}. We will focus on the first,
where the integrals, essentially, are replaced by sums.

Consider the CAA system \eref{eq:game_model1}-\eref{eq:game_model4} with discrete
variables as given in \sref{sec:discrete_caa}. Define the following recursive sequence of
belief sets:
\begin{equation}
    \bset_k \eqdef \left\{ T(\pi, y) : y \in \mathcal{Y}, \pi \in \bset_{k-1} \right\},
    \label{eq:bset_def_discrete}
\end{equation}
initiated with $\bset_0 = \{\pi_0\}$. Then, from \thrmref{thrm:inverse_filter}, it follows
that the optimal inverse filter has the following finite-dimensional form:
\begin{corollary}
    For a discrete CAA system (defined in \sref{sec:discrete_caa}), the optimal inverse
    filter \eref{eq:alpha_recursion} takes the form
    \begin{equation}
        \fv{k+1}(\pi) = \frac{\rpolicy_{\pi, \action_{k+1}} \sum_{\bar{\pi} \in \bset_k}
        B_{x_{k+1}, y_{\bar{\pi}, \pi}} \fv{k}(\bar{\pi})}{\sum_{\pi \in \bset_{k+1}}
    \rpolicy_{\pi, \action_{k+1}} \sum_{\bar{\pi} \in \bset_k} B_{x_{k+1}, y_{\bar{\pi},
\pi}} \fv{k}(\bar{\pi})},
        \label{eq:alpha_recursion_discrete}
    \end{equation}
    where $\bset_k$ is defined in \eref{eq:bset_def_discrete}. The conditional mean
    estimate of the adversary's belief is 
    \begin{equation}
        \E{\pi_{k+1} | \action_{1:k+1}, x_{0:k+1}} = \sum_{\pi \in \Pi_{k+1}} \pi \fv{{k+1}}(\pi).
        \label{eq:cond_mean_discrete}
    \end{equation}
    \label{cor:inverse_filter_discrete}
\end{corollary}

\begin{remark}

    It should be noted that $B_{x, \varnothing} = 0$ -- which happens if there is no
    observation $y_{\bar{\pi}, \pi}$ that maps $\bar{\pi}$ to $\pi$ via the filter $T$.

\end{remark}

\section{Optimal Fixed-Interval Smoother for Estimating Beliefs in CAA Systems}
\label{sec:inv_smooth}

We are now in a position to derive the main theoretical result of this paper: Given knowledge
of our state sequence $x_{0:N}$ and recorded actions of the adversary $a_{1:N}$, what can
be said about the corresponding (for us, unobserved) sequence of beliefs $\pi_{1:N}$ that
were held by the adversary? More specifically, we aim to determine the conditional
distribution of the belief at time $k$ given measurements up to time $N \geq k$ (i.e.,
\pref{pr:inverse_filter_caa}b). This is a well-studied problem for partially observed
dynamical models and the task is generally referred to as \emph{smoothing}
\cite{krishnamurthy_partially_2016, cappe_inference_2005}.

\begin{figure}[t!] 
    \centering

    \includegraphics{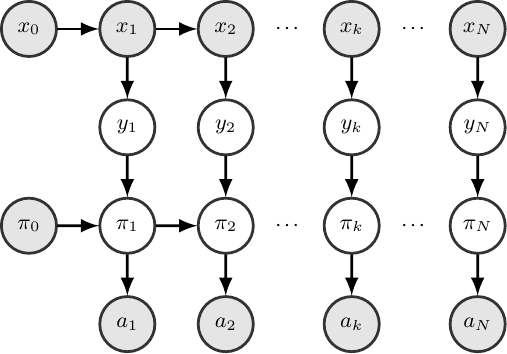}

    \caption{The graphical model underlying the inverse filtering problem
        for CAA systems -- i.e., \eref{eq:game_model1}-\eref{eq:game_model4}. The shaded nodes are
        known to us, and the white nodes are known only to the adversary. Our aim is to
        estimate the adversary's beliefs $\pi_1, \dots, \pi_N$.}

    \label{fig:inverse_filt_graph}
\end{figure}

\subsection{Optimal Smoother for CAA Systems}

We focus on determining the \emph{fixed-interval smoother} in which the number of
measurements $N$ is kept fixed and the aim is to determine the conditional distribution of
$\pi_k$ for values of $k$ between 1 and $N$ -- the underlying graphical model is
illustrated in \fref{fig:inverse_filt_graph}. From this result, it is easy to derive,
e.g., the \emph{fixed-lag smoother} (where one tries to estimate the adversary's belief
some fixed number of time-steps in the past for an ever-increasing number of measurements
$N$).

Our key result is the following theorem that provides a recursive algorithm for computing
the (fixed-interval) smoothing distribution
\begin{equation}
    \fv{{k|N}}(\pi) \eqdef p(\pi_k = \pi | \action_{1:N}, x_{0:N}),
\end{equation}
where $N$ is fixed and $1 \leq k \leq N$.

\begin{theorem}
    The smoothing distribution $\fv{k|N}(\pi)$ satisfies 
    \begin{equation}
        \fv{{k|N}}(\pi) = \frac{\bv{k|N}(\pi) \fv{k}(\pi)}{\int_{\bset}
        \bv{k|N}(\pi) \fv{k}(\pi) d\pi},
        \label{eq:fv_kN_def}
    \end{equation}
    where $\fv{k}(\pi)$ is the optimal inverse filter \eref{eq:alpha_recursion} -- or, the
    \emph{forward variable} -- and the \emph{backward variable} $\bv{k|N}(\pi)$ can be
    computed recursively via
    \begin{equation}
        \bv{k|N}(\pi) = \int_{\bset} \rpolicy_{z, \action_{k+1}} P_{x_k, x_{k+1}} B_{x_k, y_{\pi,
        z}} \bv{k+1|N}(z) d z,
        \label{eq:bv_recursion}
    \end{equation}
    initialized by
        $\bv{N|N}(\pi) = 1,$
    for all $\pi \in \bset$.

    \label{thrm:general_smoother}
\end{theorem}
Note that, as in \thrmref{thrm:inverse_filter}, $y_{\pi, z}$ is the observation such that $z = T(\pi,
y)$, where $T$ is the adversary's filter \eref{eq:game_model3}, and the smoothed
conditional-mean estimate is
\begin{equation}
    \E{ \pi_k | \action_{1:N}, x_{0:N} } = \int_{\bset} \pi \fv{{k|N}}(\pi) d\pi.
    \label{eq:cme_smoother}
\end{equation}

\begin{remark}

    We do not refer to this as an ``optimal inverse smoother'', but rather as an ``optimal
    smoother for the inverse filtering problem for CAA systems''. It is the adversary's
    \emph{filter} we try to invert (by using a smoother); the adversary is not employing
    a smoother.

\end{remark}

\subsection{Optimal Smoother for Discrete CAA Systems}

We now show how \thrmref{thrm:general_smoother} can be applied to discrete CAA systems.
Even though the state and observation spaces are discrete, the corresponding backward
variable $\bv{k|N}(\pi)$ does not allow for a finite-dimensional characterization.
Fortunately, however, in order to compute the smoother $\fv{k|N}(\pi)$, it is sufficient
to \emph{evaluate} the backward variable in a finite number of points.  Below, we describe
how to obtain its values in these points recursively. 

\begin{theorem}
    For a discrete CAA system (defined in \sref{sec:discrete_caa}), the smoother
    $\fv{k|N}(\pi)$ can be evaluated via
    \begin{equation}
        \fv{{k|N}}(\pi) = \frac{\bvr{k|N}(\pi) \fv{k}(\pi)}{\sum_{z \in \bset_k}
        \bvr{k|N}(z) \fv{k}(z)},
        \label{eq:discrete_smoother_evaluation}
    \end{equation}
    where $\fv{k}(\pi)$ is the optimal inverse filter \eref{eq:alpha_recursion} and
    $\bvr{k|N}(\pi)$ is the null-extended restriction of $\bv{k|N}(\pi)$ to $\bset_k$
    (defined in \eref{eq:bset_def_discrete}):
    \begin{equation}
        \bvr{k|N}(\pi) = 
        \begin{cases}
            \bv{k|N}(\pi) &\text{ if } \pi \in \bset_k, \\
            0 &\text{ otherwise. }
        \end{cases}
        \label{eq:bvr_definition}
    \end{equation}
    The restriction of $\bv{k|N}(\pi)$ to $\bset_k$ can be computed recursively via 
    \begin{equation} 
        \bv{k|N}(\pi) = \sum_{z \in \bset_{k+1}} \rpolicy_{z, \action_{k+1}} P_{x_k, x_{k+1}} B_{x_k, y_{\pi, z}} \bv{k+1|N}(z),
        \label{eq:discrete_backward_recursion}
    \end{equation}
    for $\pi \in \bset_{k}$, initialized by
        $\bv{N|N}(\pi) = 1$
    for all $\pi \in \bset_N$.

    \label{thrm:discrete_smoother}
\end{theorem}

In summary, to evaluate the smoothing distribution $\fv{k|N}(\pi)$ for a discrete CAA
system, one
\begin{enumerate}
    \item[\emph{i)}] computes the optimal inverse filter $\fv{k}(\pi)$ via
        \eref{eq:alpha_recursion_discrete};
    \item[\emph{ii)}] computes the backward variables $\bv{k|N}(\pi)$ on the points in the
        sets $\bset_k$ via the recursion in \eref{eq:discrete_backward_recursion}, and
        uses this to obtain the null-extended restricted backward variable
        $\bvr{k|N}(\pi)$ via \eref{eq:bvr_definition};
    \item[\emph{iii)}] combines the filter $\fv{k}(\pi)$ and $\bvr{k|N}(\pi)$ using \eref{eq:discrete_smoother_evaluation}.
\end{enumerate}


\begin{figure}[t!]
\centering
\begin{subfigure}[b]{0.45\columnwidth}
    \centering

    \includegraphics{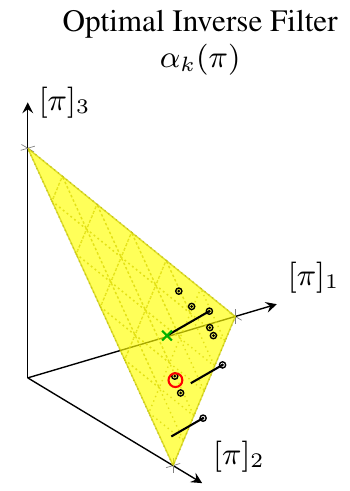}

\end{subfigure}
\begin{subfigure}[b]{0.05\columnwidth}
    $ $
\end{subfigure}
\begin{subfigure}[b]{0.45\columnwidth}

    \includegraphics{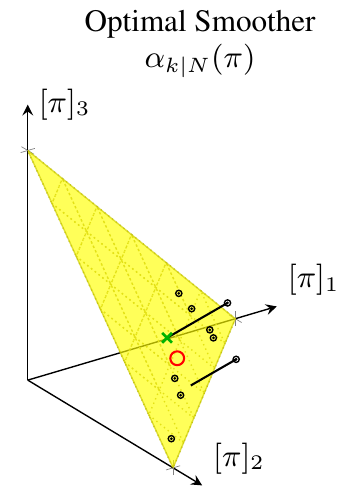}

\end{subfigure}

\caption{The figures show, left, the optimal inverse filter $\fv{k}(\pi)$, and, right, the
    smoother $\fv{k|N}(\pi)$ at time $k = 3$ and $N = 6$. The bars display the
    probability mass function (some beliefs have zero probability). The actual
    belief of the adversary (on the yellow unit simplex) is marked with a green cross, and
the \emph{conditional mean estimate} (CME) by a red circle. One should note that the
smoother's CME lies closer to the actual belief -- hence, providing a better
estimate.\looseness=-1}

\label{fig:numerical_results}

\end{figure}

\section{Numerical Experiments}
\label{sec:numerical}


In this section, we illustrate and evaluate the theoretical results in numerical
simulations. We consider a three-state system so that $\pi_k \in \Rb^3$, and the filter
$\fv{k}(\pi)$ and smoother $\fv{k|N}(\pi)$ yield \emph{probability mass functions} (pmfs)
over the 2-dimensional unit simplex. 
\indent In particular, we consider the following CAA system:
\begin{equation}
    P = \begin{bmatrix}
        0.7 & 0.2 & 0.1 \\
        0.1 & 0.4 & 0.5 \\
        0.1 & 0.1 & 0.8
    \end{bmatrix}, \quad
    B = \begin{bmatrix}
        0.3 & 0.3 & 0.4 \\
        0.1 & 0.8 & 0.1 \\
        0.1 & 0.4 & 0.5
    \end{bmatrix},
\end{equation}
with $\mathcal{A} = \{1,2\}$ and a $\rpolicy$ that yields the first action if
$[\pi_k]_1 \geq 0.5$, and the second action otherwise.

\vspace{-0.2cm}

\subsection{Illustration}

In the left plot of \fref{fig:numerical_results}, we illustrate the pmf of the optimal
inverse filter $\fv{3}(\pi) = p(\pi_3 = \pi | a_{1:3}, x_{0:3})$ computed via
\eref{eq:alpha_recursion_discrete}. We have marked the \emph{conditional mean estimate}
(CME) with a red circle, and the adversary's actual belief with a green cross. It should be
noted that the optimal inverse filter assigns zero probability to several points in the
set $\bset_3$. 

In the right plot, we illustrate the smoother $\fv{3|6}(\pi) = p(\pi_3 = \pi | a_{1:6},
x_{0:6})$ computed via \eref{eq:discrete_smoother_evaluation}. Its CME and the
adversary's actual belief are marked as before. The smoother, having access to additional
data (i.e., the actions $\action_{4:6}$ and states $x_{4:6}$), rules out one of the
potential beliefs of the adversary. Consequently, its CME is closer to the actual belief
of the adversary. 

\vspace{-0.2cm}

\subsection{Improved Accuracy in Estimating the Adversary's Beliefs}

\begin{figure}[t!]
    \centering

    \includegraphics{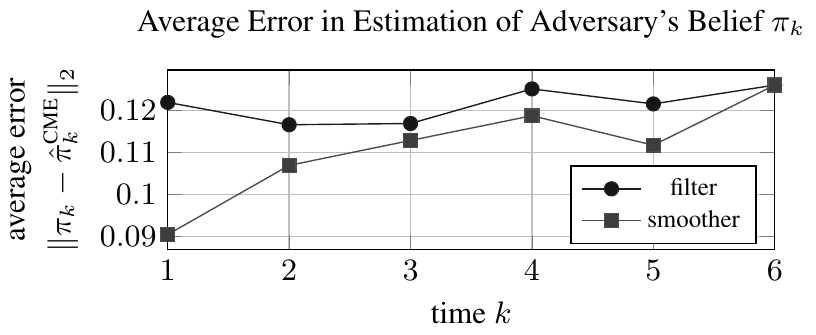}

    \caption{Average error of the \emph{conditional mean estimate} (CME) of the inverse
        filter and smoother, compared to the adversary's actual belief $\pi_k$. The
        smoother yields, on average, more accurate estimates. Note that the smoother
        and the filter coincide at the last point in the interval -- that is, $\fv{N}(\pi)
        = \fv{N|N}(\pi)$.  }

    \label{fig:avg_cme_error}
\end{figure}

Next, we compute the error between the actual belief of the adversary $\pi_k$ and the CMEs
of the optimal inverse filter \eref{eq:cond_mean_discrete} and the smoother
\eref{eq:cme_smoother} for various values of $k$ between 1 and $N = 6$. We average the
errors over 1000 realizations. The results are in \fref{fig:avg_cme_error}.

The smoother yields, on average, a lower error than the filter. Its estimate of the
adversary's actual belief is better since it can incorporate more information -- not only
measurements up to time $k$ when estimating $\pi_k$, but also those from times $k+1,
\dots, N$. It should be noted that the filter $\fv{k}(\pi)$ and smoother $\fv{k|N}(\pi)$
coincide for $k = N$, and hence, yield the same average error. 


\vspace{-0.2cm}

\section{Conclusion and Discussion}

\vspace{-0.1cm}

We have derived the optimal smoother for inverse filtering in \emph{counter-adversarial
autonomous} (CAA) systems -- the goal being to estimate an adversary's beliefs given
observed actions and knowledge of the state sequence. As expected, the smoother is more
accurate (in terms of mean-squared errors) than the optimal inverse filter because it has
access to more information, which we verified in numerical simulations. Moreover, we proposed
a finite algorithm for discrete CAA systems.

Future work includes studying the important problem of mismatched systems
(e.g., where the adversary does not have perfect knowledge of the transition
kernel $P$) and dealing with the computational concerns resulting from the
exponential growth of the sets of potential beliefs ($|\Pi_k| = Y^k$) in
discrete CAA systems using, for example, particle filters and smoothers
\cite{doucet_tutorial_2009}.



\clearpage
\balance
\bibliography{rob_references.bib,rob_references_smoother.bib}

\clearpage
\newpage
\nobalance


\section{The Complete CAA System Model}

In more generality, the CAA model \eref{eq:game_model1}-\eref{eq:game_model4} takes the
form:
\begin{align*}
    \text{us:}\quad x_k &\sim P_{x_{k-1}, x} = p(x | x_{k-1}), \;\; x_0 \sim \pi_0 \\
    \text{adversary:}\quad y_k &\sim B_{x_k, y} = p(y | x_k), \\
    \text{adversary:}\quad \pi_k &= T(\pi_{k-1}, y_k), &  \\
    \text{adversary:}\quad u_k &\sim C_{\pi_k, u} = p(u | \pi_k), & \\
    \text{us:}\quad \action_k &\sim D_{u_k, \action} = p(\action | u_k), & 
\end{align*}
where $u_k \in \mathcal{A}$ is the action taken by the adversary (according to a control policy $C$) and
$\action_k \in \mathcal{A}$ is our observation (via $D$) of it. In the paper, to
simplify, we assume we have direct access to $\rpolicy_{\pi, a} = \sum_{u \in \mathcal{A}}
D_{u,a} C_{\pi, u}$.

\section{Proof of Theorem 2}

\begin{proof}
    Consider the smoothing distribution
    \begin{align}
        \fv{k|N}(\pi) &= p(\pi_k = \pi | \action_{1:N}, x_{0:N}) \notag \\
                      &= \frac{p(\pi_k = \pi, \action_{1:N}, x_{0:N})}
                              {\int_\bset p(\pi_k = \pi', \action_{1:N}, x_{0:N}) d\pi'} \notag \\
                      &= \frac{p(\pi_k = \pi, \action_{k+1:N}, x_{k+1:N} | \action_{1:k}, x_{0:k})}
                              {\int_\bset p(\pi_k = \pi', \action_{k+1:N}, x_{k+1:N} | \action_{1:k}, x_{0:k}) d\pi'} \notag \\
                              &\hspace{-1.0cm}= \frac{p(\action_{k+1:N}, x_{k+1:N} | \pi_k = \pi, \action_{1:k}, x_{0:k}) \fv{k}(\pi)}
                              {\int_\bset p(\action_{k+1:N}, x_{k+1:N} | \pi_k = \pi', \action_{1:k}, x_{0:k}) \fv{k}(\pi') d\pi'},
    \end{align}
    where in the third equality the term $p(\action_{1:k}, x_{0:k})$ cancels in
    numerator and denominator, and in the last equality we have identified
    $p(\pi_k = \pi | \action_{1:k}, x_{0:k}) = \fv{k}(\pi)$ as the optimal
    inverse filter \eref{eq:alpha_recursion}.

    Now, note that $\action_{k+1:N}, x_{k+1:N}$ and $\action_{1:k}, x_{0:k-1}$
    are conditionally independent given $\pi_k, x_k$, so that
    \begin{align}
        \fv{k|N}(\pi) &= \frac{p(\action_{k+1:N}, x_{k+1:N} | \pi_k = \pi, x_{k}) \fv{k}(\pi)}
                              {\int_\bset p(\action_{k+1:N}, x_{k+1:N} | \pi_k
                              = \pi', x_{k}) \fv{k}(\pi') d\pi'} \notag \\
                              &= \frac{\bv{k|N}(\pi) \fv{k}(\pi)}
                              {\int_\bset \bv{k|N}(\pi') \fv{k}(\pi') d\pi'},
    \end{align}
    where we have defined
    \begin{equation}
        \bv{k|N}(\pi) \eqdef p(\action_{k+1:N}, x_{k+1:N} | \pi_k = \pi, x_{k}).
    \end{equation}
    We refer to $\bv{k|N}(\pi)$ as the the backward variable since it can be
    computed via a backward recursion. To show this, begin with 
    \begin{align}
        \bv{k|N}(\pi) &= p(\action_{k+1:N}, x_{k+1:N} | \pi_k = \pi, x_k) \notag \\
                      &= \int_\bset p(\action_{k+1:N}, x_{k+1:N}, \pi_{k+1} = z | \pi_k = \pi, x_k) dz \notag \\
                      &= \int_\bset p(\action_{k+2:N}, x_{k+2:N} | \action_{k+1}, x_{k+1}, \pi_{k+1} = z , \pi_k = \pi, x_k) \notag \\
                      & \hspace{1.5cm} \times p( a_{k+1}, x_{k+1}, \pi_{k+1} = z | \pi_k = \pi, x_k) dz.
        \label{eq:smooth_tmp0}
    \end{align}
    The first factor inside the integral equals $\bv{k+1|N}(z)$ (due to conditional
    independence), 
    and the second factor 
    \begin{align}
        p(& \action_{k+1}, x_{k+1}, \pi_{k+1} = z | \pi_k = \pi, x_k) \notag \\
                   &= \int_\mathcal{Y} p(a_{k+1}, x_{k+1}, \pi_{k+1} = z, y_{k+1} = y | \pi_k = \pi, x_k) dy
        \label{eq:smooth_tmp1}
    \end{align}
    can be factorized as follows:
    \begin{itemize}
        \item $p(\action_{k+1} | x_{k+1}, \pi_{k+1} = z, y_{k+1} = y, \pi_k = \pi, x_k)$ \\
            $= p(\action_{k+1} | \pi_{k+1} = z) = \rpolicy_{z, \action_{k+1}}$;
        \item $p(\pi_{k+1} = z | x_{k+1}, y_{k+1} = y, \pi_k = \pi, x_k)$ \\
            $= p(\pi_{k+1} = z | y_{k+1} = y, \pi_k = \pi) = \ind{z - T(\pi, y)}$,
            since the map $T$ is
            deterministic;
        \item $p(y_{k+1} = y | x_{k+1}, \pi_k = \pi, x_k)$ \\
            $= p(y_{k+1} = y | x_{k+1}) = B_{x_{k+1}, y}$;
        \item $p(x_{k+1} | \pi_k = \pi, x_k) = p(x_{k+1} | x_{k}) = P_{x_{k}, x_{k+1}}$.
    \end{itemize}
    Taken together in \eref{eq:smooth_tmp1}, we obtain
    \begin{align}
        p(\action_{k+1}, x_{k+1},& \pi_{k+1} = z | \pi_k = \pi, x_k) \notag \\
                                 &= \int_\mathcal{Y} \rpolicy_{z, \action_{k+1}}
        \ind{z - T(\pi, y)} B_{x_{k+1}, y} P_{x_{k}, x_{k+1}} dy \notag \\
        &= \rpolicy_{z, \action_{k+1}} B_{x_{k+1}, y_{\pi, z}} P_{x_{k}, x_{k+1}},
    \end{align}
    which, finally, when introduced in \eref{eq:smooth_tmp0} yields the recursion
    \begin{equation}
        \bv{k|N}(\pi) = \int_\bset \bv{k+1|N}(z) \rpolicy_{z, \action_{k+1}} B_{x_{k+1},
        y_{\pi, z}} P_{x_{k}, x_{k+1}} dz.
    \end{equation}
    Here, we have defined $y_{\pi, \bar{\pi}}$ as the $y \in \mathcal{Y}$ such that
    $\bar{\pi} = T(\pi, y)$.

    Since $\bv{N|N}(\pi) = p(\action_{N+1:N}, x_{N+1:N} | \pi_N = \pi, x_{N})
    = p(\varnothing | \pi_N = \pi, x_N) = 1$, the recursion is initialized by
    \begin{equation}
        \bv{N|N}(\pi) = 1,
    \end{equation}
    for all $\pi \in \bset$.
\end{proof}

\section{Proof of Theorem 3} 

\begin{proof}
    The theorem follows by induction. First, note that we can trivially evaluate
    $\bv{N|N}(\pi)$ for all $\pi \in \bset_N$ (since $\bv{N|N}(\pi) = 1$ for all $\pi \in
    \bset$).

    Now, suppose we can evaluate $\bv{k+1|N}(\pi)$ for all $\pi \in \bset_{k+1}$. Then,
    consider evaluating $\bv{k|N}(\pi)$ for $\pi \in \bset_k$ via \eref{eq:bv_recursion}:
    \begin{equation}
        \bv{k|N}(\pi) = \int_{\bset} \rpolicy_{z, \action_{k+1}} P_{x_k, x_{k+1}} B_{x_k,
        y_{\pi, z}} \bv{k+1|N}(z) d z.
        \label{eq:dis_sm_tmp_0}
    \end{equation}
    The factor $B_{x_k, y_{\pi, z}}$ is only non-zero for $z$:s in the set
    \begin{equation}
        \{\bar{\pi} \in \bset : \exists y \in \mathcal{Y} \text{ s.t. } \bar{\pi} = T(\pi, y) \},
        \label{eq:dis_sm_tmp_1}
    \end{equation}
    that is, for those beliefs that can be obtained from the belief $\pi$ through some
    observation $y$. By noting that we are only aiming to evaluate \eref{eq:dis_sm_tmp_0}
    for $\pi \in \bset_k$, we see that the set \eref{eq:dis_sm_tmp_1} coincides with
    the definition of $\bset_{k+1}$ given in \eref{eq:bset_def_discrete}. Hence, 
    \begin{align}
        \bv{k|N}(\pi) &= \int_{\bset} \rpolicy_{z, \action_{k+1}} P_{x_k, x_{k+1}} B_{x_k,
        y_{\pi, z}} \bv{k+1|N}(z) d z \notag \\
        &= \int_{\bset_{k+1}} \hspace{-0.4cm} \rpolicy_{z, \action_{k+1}} P_{x_k, x_{k+1}} B_{x_k, y_{\pi, z}} \bv{k+1|N}(z) d z \notag \\
        &= \sum_{z \in \bset_{k+1}} \hspace{-0.1cm} \rpolicy_{z, \action_{k+1}} P_{x_k,
    x_{k+1}} B_{x_k, y_{\pi, z}} \bv{k+1|N}(z). 
    \end{align}
    By our induction-assumption, we can evaluate $\bv{k+1|N}(z)$ for $z \in \bset_{k+1}$.
    This demonstrates that we have access to $\bv{k|N}(\pi)$ for $\pi \in \bset_k$.

    Observe that when evaluating the smoother $\fv{k|N}(\pi)$ via \eref{eq:fv_kN_def}, the
    inverse filter $\fv{k}(\pi)$ yields zero for any $\pi \not\in \bset_k$. Hence, the
    value of $\bv{k|N}(\pi)$ for $\pi \not\in \bset_k$ is irrelevant (since it multiplies
    $\fv{k}(\pi)$ which is zero there). We can thus, for example, use the null-extended
    restriction of $\bv{k|N}(\pi)$ to $\bset_k$:
    \begin{equation}
        \frac{\bv{k|N}(\pi) \fv{k}(\pi)}{\int_{\bset}
        \bv{k|N}(\pi) \fv{k}(\pi) d\pi} =  
        \frac{\bvr{k|N}(\pi) \fv{k}(\pi)}{\sum_{z \in \bset_k}
            \bvr{k|N}(z) \fv{k}(z)},
    \end{equation}
    which we have access to (per the above induction proof).
\end{proof}

\section{Details on CAA System \\ Used in Numerical Simulations}

In the numerical experiments, we use the following CAA system: 
\begin{equation}
    P = \begin{bmatrix}
        0.7 & 0.2 & 0.1 \\
        0.1 & 0.4 & 0.5 \\
        0.1 & 0.1 & 0.8
    \end{bmatrix}, \quad
    B = \begin{bmatrix}
        0.3 & 0.3 & 0.4 \\
        0.1 & 0.8 & 0.1 \\
        0.1 & 0.4 & 0.5
    \end{bmatrix},
\end{equation}
and $\mathcal{A} = \{1, 2\}$ with the $\rpolicy$ induced by

\begin{figure}[h!] 
\centering
\includegraphics{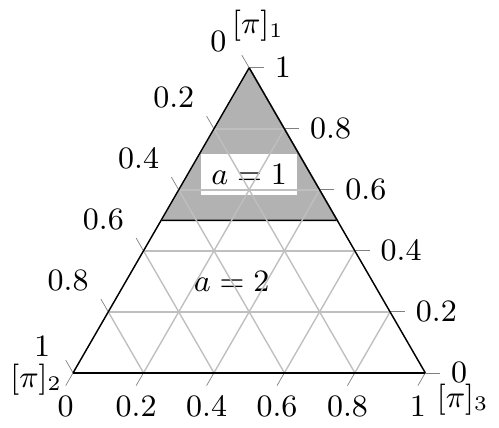}
\end{figure}
\noindent i.e.,
\begin{equation}
    \rpolicy_{\pi, a = 1} = 
    \begin{cases}
        1 & \text{ if } [\pi]_1 \geq 0.5, \\
        0 & \text{ otherwise},
    \end{cases}
\end{equation}
and
\begin{equation}
    \rpolicy_{\pi, a = 2} = 
    \begin{cases}
        1 & \text{ if } [\pi]_1 < 0.5, \\
        0 & \text{ otherwise}.
    \end{cases}
\end{equation}

\end{document}

%% file: pre_rob.tex
\usepackage{graphicx}

\usepackage[cmex10]{amsmath}
\usepackage{amsfonts}
\usepackage{amssymb}
\usepackage{accents}
\usepackage{url}

\usepackage{empheq}
\usepackage[normalem]{ulem}

\usepackage{enumerate}

\usepackage{tikz} 
\usepackage{pgfplots}
\usetikzlibrary{patterns}

\DeclareMathOperator{\diag}{diag}

\newcommand{\ind}[1]{\text{I}\{#1\}}

\newcommand{\E}[1]{\mathbb{E}\big\{#1\big\}}

\usepackage{mathtools}
\DeclarePairedDelimiter{\PrfencesHard}{[}{]}

\DeclarePairedDelimiter{\PrfencesSoft}{(}{)}
\renewcommand{\Pr}{\operatorname{Pr}\PrfencesHard}

\renewcommand{\diag}{\operatorname{diag}\PrfencesSoft}

\newcommand{\fref}[1]{Fig.~\ref{#1}}

\newcommand{\eref}[1]{(\ref{#1})}
\newcommand{\sref}[1]{Section~\ref{#1}}

\newcommand{\thrmref}[1]{Theorem~\ref{#1}}

\newcommand{\pref}[1]{Problem~\ref{#1}}

\usepackage{xcolor}
\newcommand{\comment}[1]{}  

\ifIFAC
\usepackage{enumitem} 
\else

\makeatletter%
\@ifclassloaded{IEEEconf}{
\makeatother%

}{}

\makeatletter%
\@ifclassloaded{ieeeconf}{
\makeatother%

}{}

\usepackage{amsthm}
\fi

\theoremstyle{plain} 
\newtheorem{theorem}{Theorem}

\newtheorem{corollary}{Corollary}

\theoremstyle{definition} 

\newtheorem{problem}{Problem}

\theoremstyle{remark} 
\newtheorem{remark}{Remark}

\newcommand{\Rb}{{\mathbb R}}

\usepackage{dsfont}
\newcommand{\ones}{\mathds{1}}

\newcommand{\eqdef}{\overset{\text{def.}}{=}}  

\tikzset{square arrow/.style={to path={-- ++(0,-.27) -| (\tikztotarget)}}}

\usepackage{subcaption}
